\documentclass[superscriptaddress,twocolumn,showpacs,prb]{revtex4-1}
\usepackage[utf8]{inputenc} 
\usepackage{mathtools}
\usepackage{amsmath}
\usepackage{braket}
\usepackage{graphicx}
\usepackage{pgfplots}
\pgfplotsset{compat=1.15}
\usepackage{csquotes}
\usepackage{hhline}
\usepackage{tabularx}
\usepackage{subcaption}
\usepackage{amssymb}

\usepackage{dcolumn}
\usepackage{tabularx}
\usepackage[colorlinks=true,linkcolor=blue,citecolor=blue,urlcolor=blue]{hyperref}
\usepackage{braket}
\usepackage{float}
\usepackage{listings}
\usepackage{color}
\usepackage{subcaption}
\usepackage{placeins}
\definecolor{mygreen}{rgb}{0,0.6,0}
\definecolor{mygray}{rgb}{0.5,0.5,0.5}
\definecolor{mymauve}{rgb}{0.58,0,0.82}

\newcolumntype{C}{>{\centering\arraybackslash}X}

\begin{document}

\title{Efficient Verification of Boson Sampling Using a Quantum Computer}

\author{Sritam Kumar Satpathy}
\email{sritamkumar04@gmail.com}
\affiliation{Department of Physics\\Indian Institute of Science Education and Research, Berhampur}
\author{Vallabh Vibhu}
\email{vallabhvibhu01@gmail.com}
\affiliation{Department of Physics\\Indian Institute of Science Education and Research, Berhampur}
\author{Sudev Pradhan}
\email{sudev18@iiserbpr.ac.in}
\affiliation{Department of Physics\\Indian Institute of Science Education and Research, Berhampur}
\author{Bikash K. Behera}
\email{bikas.riki@gmail.com}
\affiliation{Bikash's Quantum (OPC) Private Limited, \\Balindi, Mohanpur, 741246, Nadia, West Bengal, India}
\affiliation{Department of Physical Sciences,\\ Indian Institute of Science Education and Research Kolkata, Mohanpur 741246, West Bengal, India}
\author{Prasanta K. Panigrahi}
\email{pprasanta@iiserkol.ac.in}
\affiliation{Department of Physical Sciences,\\ Indian Institute of Science Education and Research Kolkata, Mohanpur 741246, West Bengal, India}

\begin{abstract}
Boson sampling is a sub-universal model used to show quantum speed-up. However, the methods of validation to prove quantum speedup are not robust and accurate. All verification methods involve additional or little studied assumptions. Here, we use the protocols given in the paper [arXiv:2006.03520] to construct a boson sampling experiment using discrete quantum states on IBM quantum computer and verify the fidelity of the output states using heterodyne detection. We demonstrate the protocols for single mode fidelity estimation, multi mode fidelity estimation and a verification protocol using IBMQ ``athens" chip. Moreover, we illustrate the use of this verification protocol in the quantum key distribution (QKD) process for estimating the fidelity of different types of encoding-decoding basis. This shows that the verification protocols can be used to enable efficient and reliable certification of highly entangled multi-particle states. 
\end{abstract}
 

\maketitle
\section{Introduction}

Quantum advantage or quantum speed-up \cite{bs_quantadva,bs_quantadva2,bs_quantspeed} refers to the ability of quantum computers to solve computation problems at a faster rate than a classical computer. With usage of Shor’s factoring algorithm \cite{bs_shor} such ability can be demonstrated through a universal quantum computer. However, the complexity required to build a universal quantum computer retaining all the advantages granted by quantum computation makes realization of such a machine impossible with current technology \cite{bs_near}. Hence, different sub-universal models have been realized \cite{bs_mod,bs_mod1,bs_mod2,bs_mod3,bs_mod4} which retain the computational powers and advantages of the proposed universal model to a certain degree. These sub-universal models are assigned with specific problems, which are deemed difficult to compute using classical computer, even with certain approximation or assumptions. Quantum supremacy\cite{bs_mod3} can be obtained when a quantum computer becomes powerful enough to accomplish many calculations that a classical computer cannot. The demonstration of quantum speed-up \cite{bs_demspeed} involves quantum devices that efficiently solve several sampling tasks which are marked as difficult to simulate using classical computer even under reasonable theoretical assumptions. The sub-universal models aim to demonstrate the virtues of quantum advantage through simulation of several sampling problems \cite{bs_submodel, bs_submodel1}, which involve the task to pick random samples from an ideal probability distribution model.\\

Boson sampling \cite{bs_bs} is a simplified sub-universal model designed to verify quantum speed-up, which consists of three parts: (1) vacuum and single-photon Fock states, (2) a passive linear interferometer, and (3) an on-off photo-detector \cite{bs_intro}. This model uses bosons for sampling from the probability distribution of identical photons scattered by a linear interferometer \cite{bs_lininter} using a finite number of measurements. It acts as an important demonstration of passive linear optics outperforming classical computers \cite{bs_intro}, since this sampling task has been shown as difficult to compute. Boson sampling is the simplified model for quantum computing which is the best way to implement for the first-ever post-classical-quantum computer, a non-universal quantum computer introduced by Aaronson and Arkipov \cite{bs_classcomplex}. The boson sampling model is believed to implement computing tasks by creating an environment for the photons to be introduced in an interferometer for a given period. It is based on photon loss and demonstrates that boson sampling with a few photon losses can increase the sampling rate, it can also be used with a modified input state to generate molecular vibronic spectra \cite{bs_molvib}. Use of boson sampling can be further extended towards cryptographic applications \cite{bs_cry,bs_cry1,bs_cry2}, construction of quantum-inspired classical computational algorithms which are used to estimate certain matrix permanents \cite{bs_matper}, and build superconducting resonator network sampling devices \cite{bs_res} to analyse decoherence and interferometric sensitivity.

Most of the works on boson sampling have been performed using photons in an optical setup where various aspects about the demonstration \cite{bs_photondem,bs_photodem1}, methods of validation \cite{bs_photonvalid,bs_photonvalid1} and the applications in molecular vibronic spectra, molecular docking \cite{bs_moldock} etc., are discussed using it's variant types such as scattershot \cite{bs_photonvalid1,bs_scatter}, Gaussian \cite{bs_gaussian,bs_gaussian1}, classically stimulable tasks \cite{bs_class,bs_class1} or alternate photonic pathways \cite{bs_path,bs_path1}. Several methods have been tested throughout this problem in order to find the most optimum method to demonstrate and validate the procedure while accepting minimum errors, which strongly supports the idea of quantum speed-up. For verification of sub-universal models like boson sampling, a protocol that verifies quantum speed-up experiment should ensure that for the accuracy of the experimental probability distribution relative to  ideal probability distribution \cite{bs_quantadva2} and should guarantee in terms of  total variation distance the proximity between the experimental probability and the ideal probability distributions. But verification of each challenging is difficult due to sampling occurring over an exponentially huge sample space with an approximately uniform probability distribution. However, any efficient non interactive verification of present quantum speed-up experiments with a verifier which is  restricted to classical computations requires additional cryptographic \cite{bs_crypto} assumptions which are either little studied \cite{bs_stud,bs_stud1,bs_stud2} or involves extra indirect assumptions \cite{bs_ass,bs_ass1,bs_ass2}. Weaker but more resource efficient verification methods with a classical verifier called validation \cite{bs_valid}  consisting of only specific properties of the experimental probability distribution to be tested. Hence, validation depends on making extra undesirable assumptions about the internal function of the quantum device \cite{bs_assu}.

In our procedure, we have used single-mode Gaussian measurement namely heterodyne detection \cite{bs_hetero} and three non-interactive protocols \cite{bs_proto} for the verification of discrete variable quantum state on a quantum circuit using IBM quantum computer. Protocol 1 uses heterodyne detection to estimate the fidelity of an unknown and arbitrary single-mode discrete variable quantum state. Protocol 2 uses heterodyne detection for obtaining a reliable estimate of fidelity witness by using single-mode fidelity estimation protocol as a subroutine for a large class of multi-mode discrete variable quantum states. Finally, protocol 3 is used to verify the output states of boson sampling experiment estimation protocol for multi-mode fidelity witness \cite{bs_mfid} as a subroutine. This protocol gives the total variation distance between the ideal and experimental probability distributions valid for any observable to enable a convincing demonstration of quantum speed-up. The verification protocol is enforced within the very same setup by substituting the output detectors with balanced and unbalanced heterodyne detection.

The rest of the paper is structured as follows. In sec. \ref{SecII}, we discuss single-mode fidelity estimation protocol, after then in sec. \ref{secIII}, we demonstrate the multi-mode fidelity estimation protocol. In sec. \ref{SecIV}, we present the verification of boson sampling experiment  and in sec. \ref{secV}, we discuss the applications of boson sampling in quantum cryptography and quantum key distribution. We then finally conclude in sec. \ref{secVI} by summarizing the results and discussing future directions of the work.

\begin{figure}[h]
\includegraphics[width=0.53\textwidth]{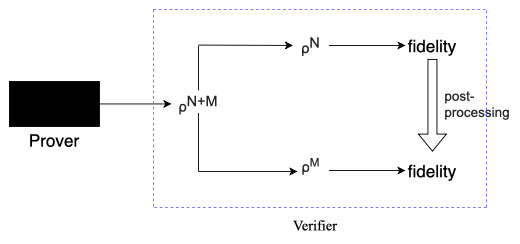}
\caption{\textbf{A pictorial representation of our protocols.} The black box represent the prover that provides $N+M$ copies of some particular quantum state, which is in the form ${\rho^{\otimes{N+M}}}$. The blue dashed rectangle represents the verifier first measures $N$ copies of the state ${\rho^{\otimes{N+M}}}$ with heterodyne detection and computes the fidelity. Thereafter, it decides whether the states that are remaining i.e. ${\rho^{\otimes{M}}}$ would be close enough to the target state by using variety of post processing techniques.}
\label{prover}
\end{figure}

\section{single-Mode  Fidelity Estimation \label{SecII}} 
\subsection{Quantum Circuit and it's Explanation}

We provide protocol-1, a method for calculating the fidelity of the discrete variable for the single mode quantum state. In protocol-1, we construct two quantum circuits containing two qubits for taking the measurement. In the first circuit shown in fig. \ref{singlemode1}, we apply a $U$ gate on the first qubit for making a superposition state whose parameters are ${\theta={\pi/2}}$, ${\phi={\pi/2}}$ and ${\lambda={\pi/2}}$ and in another Fig. \ref{singlemode4} we apply a $NOT$ gate to make the initial state as ${\ket{1}}$. In the second qubit of both the circuits, we apply a $NOT$ gate as a control to apply a controlled $U3$ gate. The parameters of controlled $U3$ gate can be changed between balanced and unbalanced heterodyne detection \cite{bs_hetero}. For ${\zeta=0}$ the parameters of controlled $U3$ gate are ${\theta=0}$, ${\phi=0}$ and ${\lambda=0}$, which is used for balanced heterodyne detection and for ${\zeta={\pi/2}}$ the parameters are ${\theta={\pi/2}}$, ${\phi=0}$ and ${\lambda=0}$, which is used for unbalanced heterodyne detection. We have used the values of $\zeta=0$ and $\zeta=\pi/2$ in both the circuits. The measurement boxes are used in the end to obtain the outputs.

For a core state, let ${\ket{C}}$=${\sum_{n=0}^{C-1}{C_n}}{\ket{n}}$ for all $C{\in{\mathbb{N^*}}}$, and also assume $N$,$M$ ${\in{\mathbb{N^*}}}$ with ${\rho^{\otimes{N+M}}}$ have $N+M$ copies for a previously unobserved single-mode quantum state ${\rho}$.  

\begin{itemize}

\item $1$. Choose $N$=$M$=5 and $C$=2, for all $N$,$M$ ${\in{\mathbb{N^*}}}$ and $C{\in{\mathbb{N^*}}}$.

\item $2$. Calculate the fidelity of five copies of ${\rho}$ using the formula $F{(a, b)}$=$\mathrm{Tr}({\sqrt{\sqrt{a}{b}{\sqrt{a}}}}$), where ${a}$ is the experimental density matrix and ${b}$ is the theoretical density matrix.

\item $3$. Calculate the mean and standard deviation of $F_1{(a,b)}$...... $F_5{(a,b)}$.

\item $4$. Plot the graph of fidelity verses number of copies of the state.

\item $5$. Follow the above same steps for the remaining five copies i.e., $M$=5. 
\end{itemize}

The measurement is done with two different conditions as mentioned above;  
(i) Initial state : ${\ket{1}}$ with ${\zeta=0}$ and ${\zeta={\pi/2}}$.
(ii) Initial state: ${\alpha{\ket{0}}}$ + ${\beta{\ket{1}}}$ with ${\zeta=0}$ and ${\zeta={\pi/2}}$.

These calculations are done using X, Y and Z measurement basis with IBMQ athens chip. We take ten different observations with each X-basis, Y-basis and Z-basis measurement for $N+M$=10. We then calculate  the experimental density matrix of the protocol-1 using the formula.

\begin{equation}
{a}=\frac{1}{2}\left[ {\begin{array}{cc}
1+\langle Z\rangle & \langle X\rangle -i\langle Y\rangle\\
\langle X\rangle + i \langle Y\rangle  & 1- \langle Z\rangle\\
\end{array} } \right]
\end{equation}

and the theoretical density matrix using the formula $b$= ${\ket{\sigma}} {\bra{\sigma}}$, where ${\sigma}$ is the final state of the first qubit before measurement. 

\subsection{Results }

\textbf {Initial state ${\ket{1}}$}:
\begin{itemize}
 \item[(i)] For ${\zeta=0}$, the fidelity is 0.9343 with standard deviation 0.0108 for the first five copies of ${\rho}$. With the same condition the fidelity is 0.9422 with standard deviation 0.0138 for the rest five copies. 
 \begin{figure}[H]
    \centering
\begin{subfigure}{.5\textwidth}
    \includegraphics[width=\linewidth]{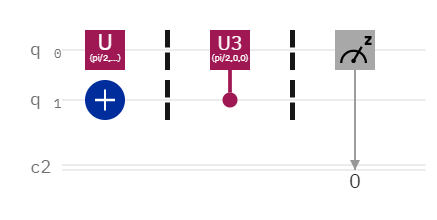}
    \caption{ For $\zeta=\pi/2$ and ${\zeta=0}$}
    \label{singlemode1}
\end{subfigure}\hfill
\begin{subfigure}{.5\textwidth}
     \includegraphics[width=\linewidth]{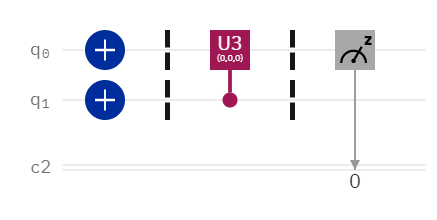}
    \caption{ For ${\zeta={\pi/2}}$ and $\zeta=0$ }
    \label{singlemode4}
\end{subfigure}\hfill
\caption{Above fig. \ref{singlemode1}, are the circuit diagrams for single-mode fidelity estimation with initial condition ${\alpha{\ket{0}}}+{\beta{\ket{1}}}$ with ${\zeta} = {\pi/2}$ and ${\zeta}=0$. Fig. \ref{singlemode4} are the circuit diagram for initial condition ${\ket{1}}$ with ${\zeta}={\pi/2}$ and ${\zeta}=0$. }
\label{singlemode}
\end{figure} 

 \begin{figure}[H]
    \centering
\begin{subfigure}{.4\textwidth}
    \includegraphics[width=\linewidth]{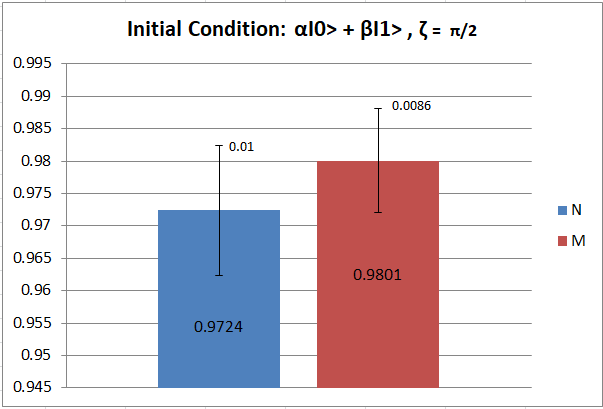}
    \caption{ For $\zeta=\pi/2$}
    \label{singlemodegraph1}
\end{subfigure}\hfill
\begin{subfigure}{.4\textwidth}
     \includegraphics[width=\linewidth]{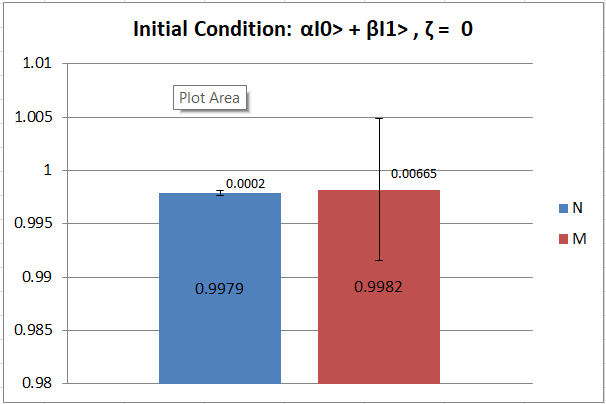}
    \caption{ For $\zeta=0$}
    \label{singlemodegraph2}
\end{subfigure}\hfill

\begin{subfigure}{.4\textwidth}
    \includegraphics[width=\linewidth]{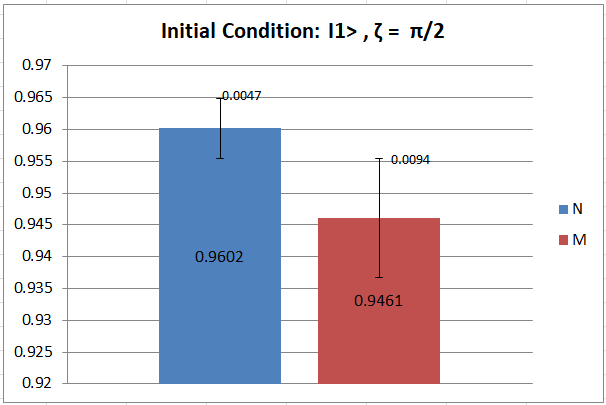}
    \caption{ For $\zeta=\pi/2$}
    \label{singlemodegraph3}
\end{subfigure}\hfill
\begin{subfigure}{.4\textwidth}
     \includegraphics[width=\linewidth]{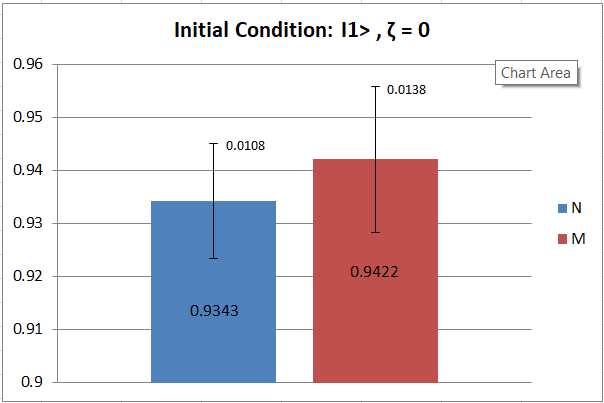}
    \caption{ For $\zeta=0$}
    \label{singlemodegraph4}
\end{subfigure}\hfill

\caption{Above figs. \ref{singlemodegraph1}, \ref{singlemodegraph2} are the graph of fidelity vs number of different copies of ${\ket{\rho}}$ for single-mode fidelity estimation with initial condition ${\alpha{\ket{0}}}+{\beta{\ket{1}}}$ with ${\zeta} = {\pi/2}$ and ${\zeta}=0$ and Fig. \ref{singlemodegraph3} and Fig. \ref{singlemodegraph4} are the graphs for initial condition ${\ket{1}}$ with ${\zeta}={\pi/2}$ and ${\zeta}=0$.}
\label{singlemode}
\end{figure} 
 \item[(ii)] For ${\zeta={\pi/2}}$, the fidelity is 0.9602 with standard deviation 0.0047 for the first five copies of ${\rho}$. With the same steps the fidelity comes out to be 0.9461 with standard deviation 0.0094 for rest five copies. 
 \end{itemize}

 \textbf{Initial State ${\alpha{\ket{0}}} + {\beta{\ket{1}}}$}:
 \begin{itemize}
 
 \item [(i)] For ${\zeta=0}$, in Fig. \ref{singlemodegraph2} the fidelity becomes 0.9979 with a standard deviation of 0.0002 for the first five copies of ${\rho}$. For rest of the five copies, the fidelity becomes 0.9982 with a standard deviation of 0.00665.
 
 \item[(ii)] For ${\zeta={\pi/2}}$, in Fig. \ref{singlemodegraph1} the fidelity is 0.9724 with a standard deviation of 0.01 for the first five copies of ${\rho}$. For rest of the five copies, the fidelity becomes 0.9801 with a standard deviation of 0.0086.
 
\end{itemize}
 Hence, Protocol-1 provides us with a reliable method with greater efficiency for evaluating the fidelity of an unknown discrete variable quantum state by comparing against ideal pure quantum state.

\section{Multi-Mode fidelity estimation \label{secIII}} 

\begin{figure}[H]
    \centering
\begin{subfigure}{.5\textwidth}
    \includegraphics[width=\linewidth]{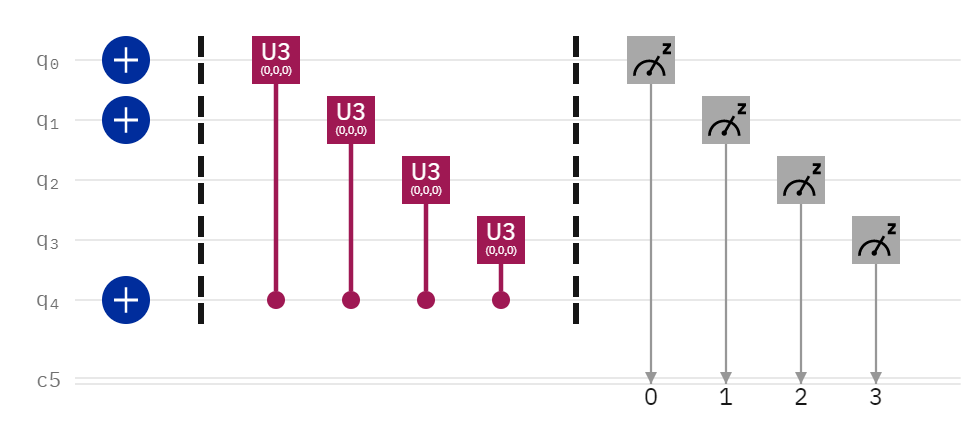}
    \caption{ For $\zeta=0$ and $\zeta=\pi/2$}
    \label{mulmode1}
\end{subfigure}\hfill
\begin{subfigure}{.5\textwidth}
     \includegraphics[width=\linewidth]{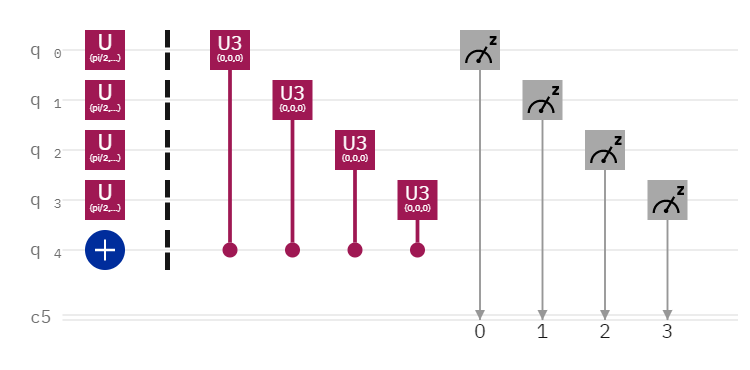}
    \caption{ For $\zeta=0$ and $\zeta=\pi/2$}
    \label{mulmode2}
\end{subfigure}\hfill

\caption{The above Fig. \ref{mulmode1} is the circuit diagram for multi-mode fidelity estimation with initial condition $\Ket{1100}$ and $\zeta = 0$ and Fig. \ref{mulmode2} is the circuit diagram for multi-mode fidelity estimation with initial condition ${\alpha{\ket{0}}}+{\beta{\ket{1}}}$ and $\zeta=0$.}
\label{multimode}
\end{figure}

This section uses single-mode fidelity estimation protocol as a subroutine to extend it to the multi-mode case. We construct a quantum circuit as given in fig. \ref{multimode} for multi-mode fidelity estimation. The multi-mode case is constructed based on the closeness between the state of all subsystems of a multi-mode quantum state ${\rho}$ and the tensor product for all the single-mode pure states present in the system. Based on the above observation, we can estimate a tight fidelity witness on the fidelity of entire system.\\

Let $\rho$ be a state over m subsystems. ${\forall}$ $i \in\{{1,.....,m}\}$ , $\rho_{i}$ =  ${Tr_{\{1,.....,m\}\backslash\{i\}}}$ $({\rho})$. Let ${\ket{\sigma_1}}$, ......., ${\ket{\sigma_m}}$ be pure states. For all $i \in\{{1,.....,m}\}$, we can write, $W_{\psi}= 1-(1-\sum_{i=1}^{m}F({a_i, b_i}))$ where $a_i$=${\ket{\rho_i}}$ ${\bra{\rho_i}}$ and $b_i$= ${\ket{\sigma_i}}$ ${\bra{\sigma_i}}$. Here, ${W_{\psi}}$ is the fidelity witness, ${F({a_i, b_i})}$ is the fidelity of $i^{th}$ qubit and ${\ket{\sigma}}$= ${\ket{\sigma_1}}$ ${\otimes}$......${\otimes}$ ${\ket{\sigma_m}}$. Hence, ${W_{\psi}}$ ${\leqslant}$ ${F({a, b})}$. 

This eq. shows that ${W_{\psi}}$ is a tight lower bound i.e. the greatest lower bound on the fidelity of the entire system. 

Let $({c_1,c_2,....c_m}) \in{\mathbb{N^*}}$, and $\ket{C_i} = \sum_{n=0}^{c_{i-1}}{c_i}{\ket{n}}$ be a core state, for all $i \in\{{1,.....,m}\}$. Let ${\rho^{\otimes{N+M}}}$ be $N+M$ the copies of an unknown m-mode quantum state ${\rho}$.

\begin{itemize}
    \item $1.$ Choose m=4, $N=1$, $C$=2 and $M=1$ for all $N$, $M$ and $C$ ${\in{\mathbb{N^*}}}$ and measure $N$ copy of ${\rho}$ with balanced heterodyne detection.
    \item $2.$ Calculate the fidelity of the system using the formula $F{(a, b)}$=$\mathrm{Tr}({\sqrt{\sqrt{a}{b}{\sqrt{a}}}})$, where ${a}$ is the experimental density matrix and ${b}$ is the theoretical density matrix.
    \item $3.$ Calculate the fidelity of individual qubits and find the fidelity witness using the  formula $W_{\psi}= 1-(1-\sum_{i=1}^{m}F_{C_i}(\rho)^N) $ where $F_{C_i}(\rho)^N$ is the fidelity of individual qubits in the system.
    \item $4.$ Follow the same above steps for measuring the remaining $M$ copies of ${\rho^{\otimes{N+M}}}$ with unbalanced heterodyne detection.
\end{itemize}

\subsection{Quantum circuit and it's explanation}

Heterodyne detection is used to create an unnormalized state for a qubit. It is performed by $U3$ gate on the first four qubits in the circuit diagram using the fifth qubit as the control one. The control $U3$ gate parameters can be changed to switch between balanced heterodyne detection ($\zeta=0$) and unbalanced heterodyne detection ($\zeta=\pi/2$). The measurement box is later used to calculate output for a given quantum circuit.\\

In our setup, we take quantized states of $\ket{0}$ and $\ket{1}$ instead of a continuous variable quantum states. Our aim is to calculate the fidelity of the quantum circuit using theoretical density matrix ($b$) and experimental density matrix ($a$), given as $F{(a,b)}$=$\mathrm{Tr}({\sqrt{\sqrt{a}{b}{\sqrt{a}}}})$. The density matrix of a quantum state $\ket{\psi} = \alpha \ket{0}+\beta \ket{1}$ is given as $\ket{\psi}\bra{\psi} = \alpha\alpha^*\ket{0}\bra{0} + \beta\beta^*\ket{1}\bra{1} +\beta\alpha^*\ket{1}\bra{0} +\alpha\beta^*\ket{0}\bra{0}$.

The measurement will be performed on the first four qubits using X-basis, Y-basis, and Z-basis based upon 256 different possible combinations. Hence, the outcomes will be in entangled states of five qubits. To calculate the experimental density matrix for the first four qubits, we trace out the fifth qubit using the formula $a_{1234}$= $\bra{1}a\Ket{1}$, where $a_{1234}$ is the traced out density matrix of the first four qubits while $a$ is the density matrix for the entire qubit. Using the same principle, we trace out the density matrix for individual qubit by tracing out the rest of the qubits. This will give us the fidelity for the collection of four qubit system as well as single qubits in the system. Using this information, we calculate the fidelity witness estimation of the first $N$ copies using the formula $W_{\psi}= 1-(1-\sum_{i=1}^{m}F_{C_i}(\rho)^N)$, where $F_{C_i}(\rho)^N$ is the fidelity of individual qubits of the system and similarly for $M$ copies.\\

In the multi-mode fidelity estimation, we first construct a normalized core state using a $NOT$ gate and then use a $U$ gate for heterodyne detection by using a controlled $U3$ gate. The measurement is done with two different cases:- (i) initial state : ${\ket{1100}}$ with ${\zeta=0}$ and ${\zeta={\pi/2}}$ and, (ii) initial state : ${\alpha_i{\ket{0}}}$ + ${\beta_i{\ket{1}}}$ with ${\zeta=0}$ and ${\zeta={\pi/2}}$. The data collected from IBMQ is used to construct experimental density matrix which is used later to calculate the fidelity of the multi-mode system. After calculating the theoretical density matrix for the given quantum state, we will use it to calculate the fidelity to verify the distance from pure states.

\subsection{Results}

\textbf{Initial state :${\ket{1100}}$ }
\begin{itemize}

\item [(i)] For ${\zeta=\pi/2}$, the fidelity comes out as 0.6983. The fidelity witness for the initial state ${\ket{1100}}$ with ${\zeta=\pi/2}$ comes out to be -0.1580.

\item [(ii)] For $\zeta=0$, the fidelity comes out as 0.6681 while the fidelity witness for ${\ket{1100}}$ with $\zeta=0$ comes out to be -1.9387.

\textbf{Initial state : ${\alpha_i{\ket{0}}}+{\beta_i{\ket{1}}}$}  

\item [(i)] $For {\zeta={\pi/2}}$, we calculate the experimental and theoretical density matrix to get the the fidelity. The fidelity comes out to be 0.7907 while fidelity witness for this state is -1.036.

\item [(ii)] For ${\zeta=0}$, the fidelity for this quantum state comes out to be 1.1004 with fidelity witness as -0.1623. The fidelity should be in the range of 0 and 1. But in this case it is greater than 1 due to noise in the chip.
\end{itemize}
Using the protocol-1 and protocol-2, we efficiently estimated the single mode fidelity and the fidelity witness for individual qubit in multi-mode case. However, the fidelity of entire system is not estimated efficiently. In ideal condition, if the fidelity of the entire system is close to 1, then the fidelity witness is also close to 1. Therefore, we need to calculate the fidelity witness for the system to efficiently estimate the fidelity of the entire system.

\section{Verification of boson sampling\label{SecIV}}

In  boson sampling verification set up n photons are fed into m subsystems of an interferometer (n$\leq$m). The main property of this system is that the number of input photons will give the same number of output photons. The set up consists of n single photon state which are delivered to an interferometer over m modes, along with vacuum state over m-n modes and measured with unbalanced heterodyne detection. In our work, we choose the state of n photons as ${\ket{1}}$ and vacuum state as ${\ket{0}}$. The computation of interferometer is done using $U$ gates over each mode and unbalanced heterodyne detection with controlled $U3$ gate. 
 
Let ${\rho^{\otimes{N+M}}}$ be $N+M$ the copies of an unknown m-mode quantum state ${\rho}$.

\begin{figure}[h]
\includegraphics[width=0.5\textwidth]{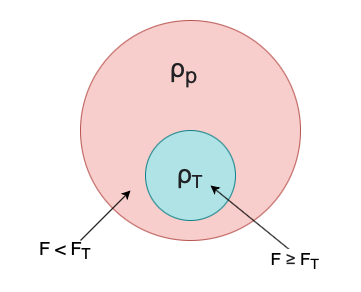}
\caption{\textbf{A pictorial representation of the fidelity and the threshold fidelity.} The verifier accepts the state ${\rho_T}$ sent by the prover, whose fidelity with the pure state is greater than or equal to the threshold fidelity (the blue region) and rejects the states ${\rho_P}$ whose fidelity with the pure state is less than the threshold fidelity (the pink region).}
\label{threshold}
\end{figure}

\begin{itemize}
\item $1.$ Choose $m=4$, $n=2$, $M=1$, $N=1$ and $C=2$ for all $N$, $M$, $C$ and $n$ ${\in{\mathbb{N^*}}}$ and measure $N$ copies ${\rho^{\otimes{N+M}}}$ with unbalanced heterodyne detection.
\item $2.$ ${\forall}$ $i \in\{{1,.....,n}\}$, compute the fidelity of $F_{i,\ket{1}}$ and ${\forall}$ $j \in\{{n+1,.....,m}\}$, compute the fidelity of $F_{j,\ket{0}}$.
\item $3.$ Compute the fidelity witness estimation using the formula ${W_{\psi}}$= ${1-\sum_{i=1}^{n}(1-F_{{i},{\ket{1}}}(\rho)^N)}$- ${\sum_{i=n+1}^{m}(1-F_{{i},{\ket{0}}}(\rho)^N)}$.
\item $4.$ Follow the same above stpes and measure the remaining $M$ copies of ${\rho^{\otimes{N+M}}}$ by using balanced heterodyne detection. 
\item $5.$ Accept, if ${F({a, b})}$ ${\geqslant}$ ${F_T}$ (threshold fidelity) fig \ref{threshold} otherwise reject the state. 
\end{itemize}

\subsection{Quantum Circuit and it's Explanation}

By taking multi-mode fidelity estimation protocol (protocol-2) as a sub-routine we develop a quantum circuit for the verification of boson sampling. In protocol-3, we construct a circuit containing five qubits among which we apply $NOT$ gate in two qubits. For making a superposition state we apply $U$ gate in the first four qubits. For our protocol we specify the different parameters of $U$ gate such as ${\theta={\pi/2}}$, ${\phi={\pi/2}}$ and ${\lambda={\pi/2}}$. Both balanced and unbalanced heterodyne detection is done by taking ${\zeta=0}$ and ${\zeta={\pi/2}}$ respectively in protocol-1 and protocol-2 using controlled $U3$ gate. Later, measurement box is used for the output of our circuit. One may guess the output state of the interferometer, but that will not be efficient.

In this protocol our main aim is to calculate the fidelity of individual qubits followed by fidelity witness. The measurement is done on the first four qubits using X-basis, Y-basis and Z-basis based upon 256 different combinations. The last qubit is taken for applying controlled $U3$ gate on different qubits. Hence we find out the experimental density matrix of the form 16 ${\times}$ 16 by taking the results of those 256 different combinations.

To find out the fidelity of first qubit we trace out the next three qubits using the formula $\bra{000}a_{234}\Ket{000}$ + $\bra{001}a_{234}\Ket{001}$ +......... $\bra{111}a_{234}\Ket{111}$ where ${a}$ is the experimental density matrix. Eight possible different combinations of the qubits ${\ket{0}}$ and ${\ket{1}}$ to be taken  into account while tracing out. After that we calculate the fidelity witness using the formula $W_{\psi}= 1-(1-\sum_{i=1}^{m}F_{C_i}(\rho)^N) $, where $F_{C_i}(\rho)^N$ is the fidelity of individual qubits in the system.

\subsection{Results}

The fidelity witness for Protocol 3 with unbalanced heterodyne detection comes out to be  -2.108 and for balanced heterodyne detection it is -2.231. \\

\begin{figure}[h]
\includegraphics[width=0.5\textwidth]{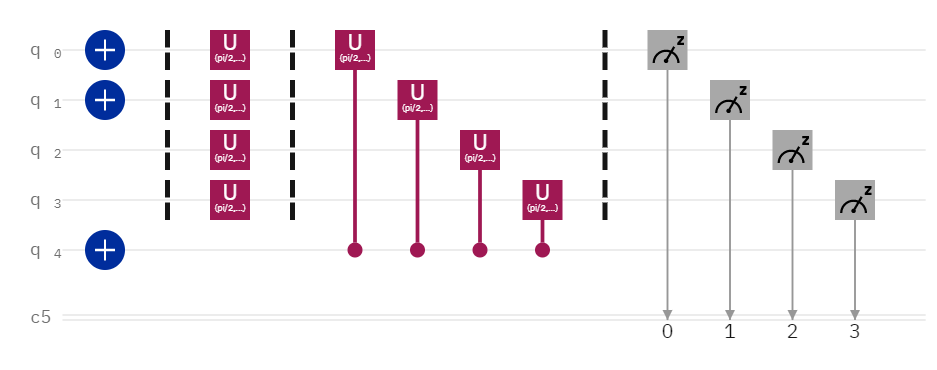}
\caption{Circuit diagram for verification of boson sampling with initial state $\Ket{1100}$ state with ${\zeta={\pi/2}}$ and ${\zeta=0}$.}
\label{verification}
\end{figure}

With discrete variable measurements within the same experimental set up, one can switch the output state of verification of boson sampling and demonstration of quantum speed-up by using the unbalanced heterodyne detection. In protocol-3 the fidelity for experimental density matrix and theoretical density matrix of the of the final output state of the circuit fig. \ref{verification} for unbalanced heterodyne detection comes out to be 0.6918 and for balanced heterodyne detection, the fidelity is 0.3113. In this paper, we choose the threshold fidelity to be 0.6. So, we reject the fidelity obtained by measuring with balanced heterodyne detection and accept the fidelity obtained by measuring with unbalanced heterodyne detection. The total variation distance and trace distance between the probability distributions corresponding to a measurement for these states ${\rho}$ and ${\sigma}$ comes out to be 0.1514 and 0.3722 respectively where ${\rho}$ is the noisy state sent by the prover and ${\sigma}$ is the target state. \\

For all ${\beta{\in[0.1514, 0.5551]}}$, ${F({a},{b})}$ ${\geqslant}$ ${1-{\beta^2}}$ that implies $\lvert\lvert {{P_\rho}-{P_\sigma}}\rvert\rvert$ ${\leqslant}$ ${\beta}$. Thus, our protocol yields a greatest lower bound on the fidelity. Also, it gives a result of the total variation distance with the target probability distribution by Fuchs- van de Graff inequality \cite{bs_inequality}, 

1-F(${a, b}$) ${\leqslant}$ D(${\rho, \sigma}$) ${\leqslant}$ ${\sqrt{1-F^2({a, b})}}$, where D(${\rho, \sigma}$) is the trace distance between the two states. In our case this inequality is as follow: 0.3082 $<$ 0.3722 $<$ 0.7221. Indeed, for the two states ${\rho}$ and ${\sigma}$, $\lvert\lvert {{P_\rho}-{P_\sigma}}\rvert\rvert$  ${\leqslant}$ D(${\rho, \sigma}$) ${\leqslant}$ ${\sqrt{1-F({a, b})}}$ \cite{bs_proto}. In our case the inequality is 0.1514 $<$ 0.3722 $<$ 0.5551. \\

So, there is a closeness between the experimental probability distribution and target probability distribution, which is hard for classical computer. This demonstrates quantum speed-up. The interferometer with unbalanced heterodyne detection is difficult to sample using classical computer. The verification of output states of boson sampling using both balance and unbalance heterodyne detection is shown efficiently.

\section{Applications of Boson sampling  \label{secV}}

Using the above concept of heterodyne detection to verify the fidelity of single-mode or multi-mode quantum states, in this section we try to implement this protocol to verify the original state of the qubit in the quantum key distribution \cite{bs_appbs,bs_appbs1} (QKD). Quantum key distribution is a type of secure communication process built on the principles of quantum mechanics. It involves sharing of an entangled pair of qubits between two users to start a secure communication channel. These shared qubits will then be used to encrypt and decrypt messages between the users so that no third party can access the information. Various quantum cryptographic protocols have been developed to ensure the safety of entangled pairs from third party access. In all these cryptographic protocols, the sender Alice encodes the information in her share of entangled qubit using her choice of basis while the receiver Bob decodes the information from his entangled pairs of qubits by measuring the outcomes using his choice of basis from a set of pre-defined basis between both the users. Then, Alice reveals all of her qubits selected basis to bob over a classical channel. Based on this information and his measurement outcomes, Bob accepts qubits having the same basis as Alice while rejects the rest. These accepted qubits can now be used to communicate between the two parties.\\

\subsection{Single-mode fidelity estimation}

\begin{figure}[H]
    \centering
\begin{subfigure}{.5\textwidth}
    \includegraphics[width=\linewidth]{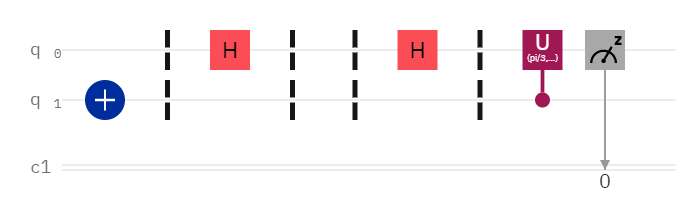}
    \caption{ \textbf{initial state ${\ket{0}}$}}
    \label{qkd single mode 1}
\end{subfigure}\hfill
\begin{subfigure}{.5\textwidth}
     \includegraphics[width=\linewidth]{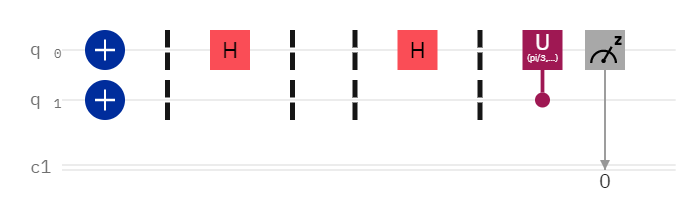}
    \caption{ \textbf{initial state ${\ket{1}}$}}
    \label{qkd single mode 2}
\end{subfigure}\hfill

\caption{Circuit diagram for different Encoding- Decoding basis with heterodyne detection for ${\zeta={\pi/2}}$ and ${\zeta=\pi/3}$}
\label{qkd single mode diagram}
\end{figure}

\begin{figure}[H]
    \centering
\begin{subfigure}{.5\textwidth}
    \includegraphics[width=\linewidth]{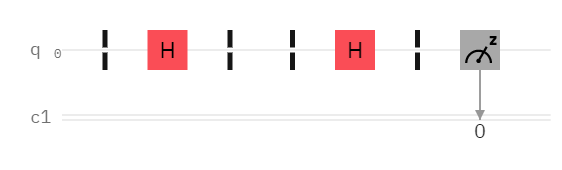}
    \caption{\textbf{initial state ${\ket{0}}$}}
    \label{qkd simple 1}
\end{subfigure}\hfill
\begin{subfigure}{.5\textwidth}
     \includegraphics[width=\linewidth]{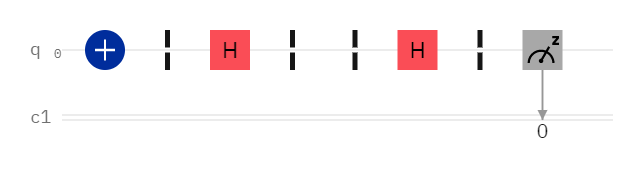}
    \caption{\textbf{initial state ${\ket{1}}$}}
    \label{qkd simple 2}
\end{subfigure}\hfill
\caption{Circuit diagram for different Encoding- Decoding Basis without heterodyne detection for single qubit fidelity}
\label{qkd simple diagram}
\end{figure}

Our aim is to introduce an alternate method for fidelity estimation so that we can clearly distinguish between the same encoding and decoding basis from a different one. In our proposed heterodyne fidelity detection model, we have designed the quantum circuit as shown in Fig. \ref{qkd single mode diagram} where we have applied an encoding basis to the  $\Ket{0}$ and $\Ket{1}$ initial  state and then a decoding basis is applied which corresponds to sender's encoding basis and receiver's decoding basis respectively while in the normal fidelity
estimation, we have not performed any heterodyne detection as shown in Fig. \ref{qkd simple diagram}. In our method of fidelity estimation, we have created an unbalanced as well as balanced heterodyne detection by applying a control
$U3$ gate using the parameter $\zeta=\pi/3$ for balanced and $\zeta=\pi/2$ parameter for unbalanced heterodyne detection, after the decoding basis which acts upon the quantum state just before measurement. Using the probability data from the real IBMQ chip ``athens" and ``qasm simulator", we have calculated the fidelity for all possible combinations  of x,y,z encoding and decoding basis. 

In the single-mode simple fidelity estimation, the difference between xz, zx, zy fidelity to the zz, yy, xx fidelity comes out to be within the range of 0.2-0.3 as shown in Table \ref{qkd1}, \ref{qkd3}. However, using our method of heterodyne detection to distinguish the same encoding, decoding states from a different one, the difference between xz, zx, zy fidelity to the zz, yy, xx fidelity comes out to be within the range of 0.5-0.6 using balanced parameter $\zeta=\pi/3$. Hence, our method for distinguishing different encoding and decoding basis gives about 200\% better distinction as compared to the normal fidelity estimation. We also give the results of fidelity estimation using unbalanced parameter $\zeta=\pi/2$. However, in this case the fidelities are not easily distinguishable therefore we need to discard the results of this parameter. We have also performed the same process using the IBM quantum simulator given in Table \ref{qkd2}, \ref{qkd4} to see the difference in the result in the absence of noise. In this case, we obtain a better fidelity estimation than using a noisy real quantum processor.

\begin{table}[]
\centering
\begin{tabular}{|c|c|c|c|c|l|}
\hline
Encoding-Decoding & ${\zeta=\frac{\pi}{3}}$ &  ${\zeta =\frac{\pi}{2}}$ & Simple fidelity\\
   Basis & & &  \\
\hline
\hline
z-z & 0.8698 & 0.7174 & 0.9985  \\ \hline
z-x & 0.2923 & 0.1566 &	0.7056  \\ \hline
z-y & 0.3185 & 0.2229 &	0.7210 \\ \hline
x-z & 0.2843 & 0.1642 &	0.7003  \\ \hline
x-x & 0.8601 & 0.7170 &	0.9977  \\ \hline
x-y & 0.7073 & 0.7125 &	0.6932  \\ \hline
y-z & 0.6958 & 0.7095 &	0.7034  \\ \hline
y-x & 0.7258 & 0.7233 &	0.7287 \\ \hline
y-y & 0.8641 & 0.7282 &	0.9979\\ \hline
\end{tabular}
\caption{Single qubit fidelity for initial state $\Ket{0}$ using real chip ``imbq athens"}
\label{qkd1}
\end{table} 

\begin{table}[]
\centering
\begin{tabular}{|c|c|c|c|c|l|}
\hline
Encoding-Decoding & ${\zeta=\frac{\pi}{3}}$ &  ${\zeta=\frac{\pi}{2}}$ & Simple fidelity\\
   Basis & & &  \\
  
\hline
\hline
z-z & 0.8662 & 0.7056  &	1  \\ \hline
z-x & 0.2608 &	1      &	0.7087  \\ \hline
z-y & 0.2647 &	1      &	0.7075 \\ \hline
x-z & 0.2532 &	1      &	0.7041  \\ \hline
x-x & 0.8686 &	0.7041 &	1  \\ \hline
x-y & 0.7074 &	0.7064 &	0.7067 \\ \hline
y-z & 0.7088 &	0.7101 &	0.7076  \\ \hline
y-x & 0.7050 &	0.7100 &	0.7137 \\ \hline
y-y & 0.8638 &	0.7089 &	1    \\ \hline
\end{tabular}
\caption{Single qubit fidelity for initial state $\Ket{0}$ using ``qasm simulator"}
\label{qkd2}
\end{table} 

\begin{table}[]
\centering
\begin{tabular}{|c|c|c|c|c|l|}
\hline
Encoding-Decoding &  ${\zeta=\frac{\pi}{3}}$ &  ${\zeta=\frac{\pi}{2}}$ & Simple fidelity\\
   Basis & & &  \\
  
\hline
\hline
z-z & 0.8393 &	0.6852 &	0.9910  \\ \hline
z-x & 0.3058 &	0.1378 &	0.7166  \\ \hline
z-y & 0.2881 &	0.1304 &	0.7183\\ \hline
x-z & 0.2924 &	0.1360 &	0.7141 \\ \hline
x-x & 0.8316 &	0.6892 &	0.9915  \\ \hline
x-y & 0.5532 &	0.6826 &	0.6917 \\ \hline
y-z & 0.7014 &	0.6950 &	0.7204  \\ \hline
y-x & 0.6950 &	0.6808 &	0.7190 \\ \hline
y-y & 0.8408 &	0.6812 &	0.9907\\ \hline
\end{tabular}
\caption{Single qubit fidelity for initial state $\Ket{1}$ using real chip ``ibmq athens".}
\label{qkd3}
\end{table} 

\begin{table}[]
\centering
\begin{tabular}{|c|c|c|c|c|l|}
\hline
Encoding-Decoding & ${\zeta=\frac{\pi}{3}}$ & ${\zeta=\frac{\pi}{2}}$ & Simple fidelity\\
   Basis & & &  \\
  
\hline
\hline
z-z & 0.8649 &	0.711 &	1  \\ \hline
z-x & 0.2646 &	1.83x$10^{-7}$ &	0.7050 \\ \hline
z-y & 0.2627 &	7.61x$10^{-7}$ &	0.7096\\ \hline
x-z & 0.2588 &	6.492x$10^{-8}$ &	0.7025 \\ \hline
x-x & 0.8660 &	0.7078 &	1  \\ \hline
x-y & 0.7120 &	0.7018 &	0.7092 \\ \hline
y-z & 0.7 &	0.6993 &	0.7053  \\ \hline
y-x & 0.7032 &	0.7021	& 0.6996 \\ \hline
y-y & 0.8666 &	0.7075 &	1\\ \hline
\end{tabular}
\caption{Single qubit fidelity for initial state $\Ket{1}$ using ``ibmq simulator".}
\label{qkd4}
\end{table}

\subsection{Multi-mode fidelity estimation}

\begin{figure}[H]
    \centering
\begin{subfigure}{.5\textwidth}
    \includegraphics[width=\linewidth]{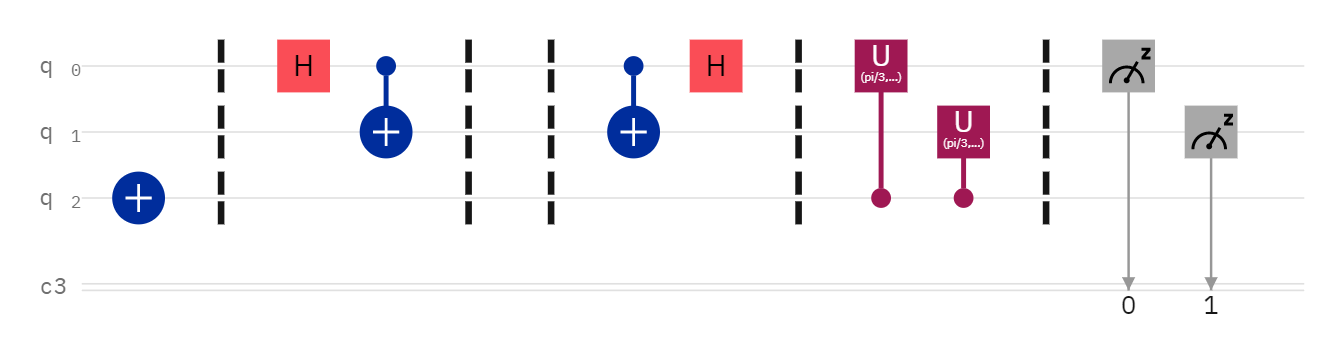}
    \caption{\textbf{ For $\beta_{00}-\beta_{00}$ encoding decoding basis}}
    \label{qkd multi di}
\end{subfigure}\hfill
\begin{subfigure}{.5\textwidth}
     \includegraphics[width=\linewidth]{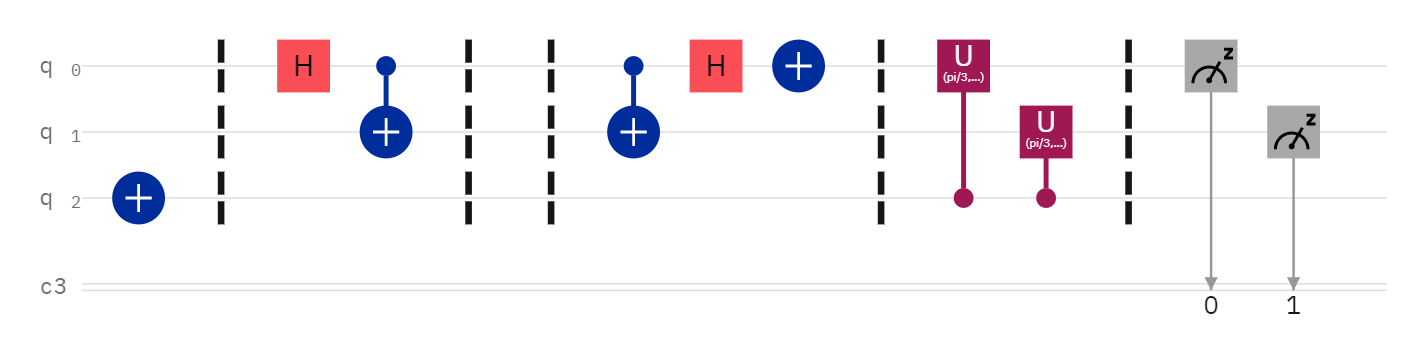}
    \caption{ \textbf{For $\beta_{00}-\beta_{10}$ encoding decoding basis}}
    \label{qkd multi di2}
\end{subfigure}\hfill

\caption{Above Fig. \ref{qkd multi di}, \ref{qkd multi di2} are the different encoding-decoding basis using heterodyne detection }
\label{qkd multimode heterodyne detection}
\end{figure}

Following the same principle as single-mode fidelity estimation, we try to construct a two qubit entangled state for multi-mode case. We designed the circuit as shown in the Fig. \ref{qkd multimode heterodyne detection} where we take the initial state as $\Ket{00}$ and encoded it using the standard bell basis $\beta_{00}$ state and decoded the state using four different bell basis given as $\beta_{00},\beta_{01},\beta_{10},\beta_{11}$. This procedure is followed in simple fidelity estimation as well as heterodyne detection. However in the simple fidelity estimation, we do not use any control $U3$ gate while in heterodyne detection, we use  control $U3$ gate using the parameter $\zeta=\pi/3$ for balanced and $\zeta=\pi/2$ parameter for unbalanced heterodyne detection after the decoding basis. 

The results of the above circuit Fig. \ref{qkd multi di}, \ref{qkd multi di2} is shown in the table \ref{qkd5}. We can clearly see from the table, the difference between various encoding-decoding basis using simple fidelity estimation and heterodyne detection. In the simple fidelity estimation, the difference between
\begin{figure}[H]
    \centering
\begin{subfigure}{.5\textwidth}
    \includegraphics[width=\linewidth]{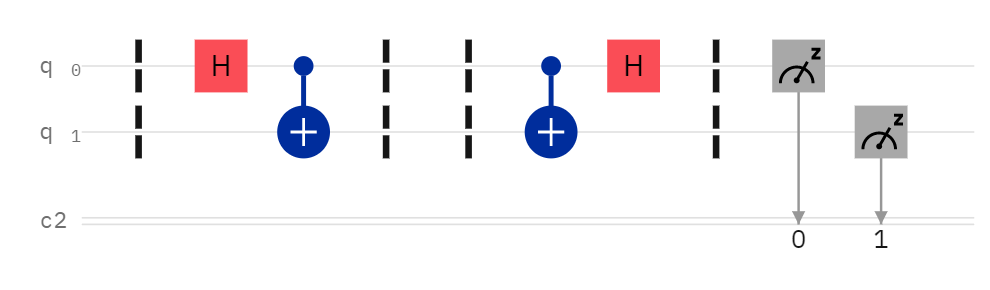}
    \caption{ \textbf{For $\beta_{00}-\beta_{00}$ encoding decoding basis}}
    \label{qkd multi di3}
\end{subfigure}\hfill
\begin{subfigure}{.5\textwidth}
     \includegraphics[width=\linewidth]{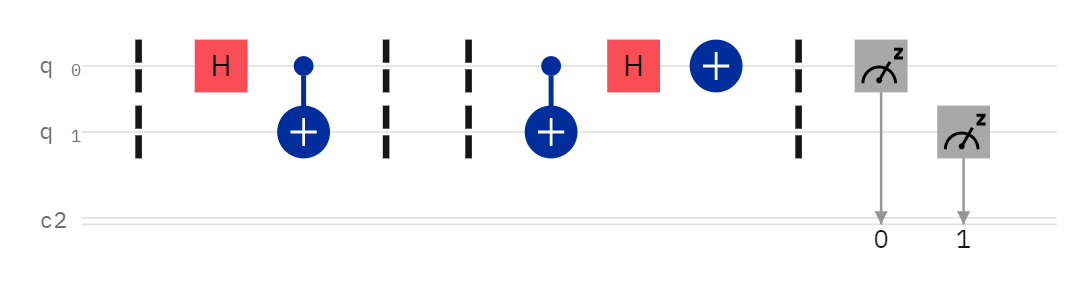}
    \caption{ \textbf{For $\beta_{00}-\beta_{10}$ encoding decoding basis}}
    \label{qkd multi di4}
\end{subfigure}\hfill

\caption{Above Fig. \ref{qkd multi di3}, \ref{qkd multi di4} are the different encoding-decoding basis without using heterodyne detection(simple fidelity) }
\label{qkd multimode simple detection}
\end{figure}$\beta_{00}-\beta_{00}$ and $\beta_{00}-\beta_{01}$ basis is 0.12 and the difference between $\beta_{00}-\beta_{10}$ and $\beta_{00}-\beta_{11}$ basis is 0.09. However, using our method of fidelity estimation, we get a difference of 0.27 for the $\beta_{00}-\beta_{00}$ and $\beta_{00}-\beta_{01}$ basis and a difference of 0.16 for the $\beta_{00}-\beta_{10}$ and $\beta_{00}-\beta_{11}$ basis. Thus, our method for distinguishing different encoding and decoding bell basis states gives about 170-230\% better distinction as compared to the simple fidelity estimation.We also give the results of fidelity estimation using unbalanced parameter $\zeta=\pi/2$. However, in this case the fidelities are not easily distinguishable therefore we need to discard the results of this parameter. We have also performed the same process given in Fig. \ref{qkd multimode simple detection} using the IBM quantum simulator given in table \ref{qkd6} to see the difference in the result in the absence of noise. In this case, we obtain a better fidelity estimation than using a noisy real quantum processor.\\

\begin{table}[H]
\centering
\begin{tabular}{|c|c|c|c|c|l|}
\hline
Encoding-Decoding & ${\zeta=\frac{\pi}{3}}$ &  ${\zeta=\frac{\pi}{2}}$ & Simple fidelity\\
   Basis & & &  \\
  
\hline
\hline
$\beta_{00}-\beta_{00}$ & 0.7458 &	0.5369 &	0.2916  \\ \hline
$\beta_{00}-\beta_{01}$ & 0.4715 &	0.5347 &	0.1650 \\ \hline
$\beta_{00}-\beta_{10}$ & 0.4467 &	0.5541 &	0.1676\\ \hline
$\beta_{00}-\beta_{11}$ & 0.2860 &	0.5058 &	0.0749 \\ \hline

\end{tabular}
\caption{multi-mode qubit fidelity for initial state $\Ket{00}$ using real chip ``ibmq athens".}
\label{qkd5}
\end{table} 

\begin{table}[H]
\centering
\begin{tabular}{|c|c|c|c|c|l|}
\hline
Encoding-Decoding & ${\zeta=\frac{\pi}{3}}$ &  ${\zeta=\frac{\pi}{2}}$ & Simple fidelity\\
   Basis & & &  \\
  
\hline
\hline
$\beta_{00}-\beta_{00}$ & 0.7550 & 0.4950 & 1	  \\ \hline
$\beta_{00}-\beta_{01}$ & 0.4369 & 0.5017 &	7.3011x$10^{-5}$ \\ \hline
$\beta_{00}-\beta_{10}$ & 0.4316 & 0.4994 &	1.0190x$10^{-4}$ \\ \hline
$\beta_{00}-\beta_{11}$ & 0.2500 & 0.5072 &	8.8430x$10^{-5}$ \\ \hline

\end{tabular}
\caption{multi-mode qubit fidelity for initial state $\Ket{00}$ using ``qasm simulator".}
\label{qkd6}
\end{table} 

\subsection{Verification of QKD}

In this part, we will set the threshold fidelity for various cases discussed above. This threshold fidelity will set the lower bound for the fidelity below which all the the various combinations of encoding-decoding basis will be rejected and we can confidently accept a quantum state as a trusted outcome. 

In the single mode fidelity estimation, the benchmark for the same encoding-decoding basis will be 0.8 for the $\zeta=\pi/3$ parameter while it is 0.9 for the simple fidelity case. Similarly for the multi mode fidelity estimation, the benchmark for the same encoding-decoding basis will be 0.7 for the $\zeta=\pi/3$ parameter while it will 0.25 for the simple fidelity case.\\

\section{Conclusion \label{secVI}}

In this work, we introduce a method for the verification of boson sampling using a quantum computer. We then explicate a new application of boson sampling experiment in the field of quantum cryptography. We derive various protocols such as protocol-1, protocol-2 and protocol-3. In protocol-1 we showed the single-mode fidelity estimation using both balanced and unbalanced heterodyne detection, which remains efficient. We use Protocol 1 as a sub-routine to set up a protocol for estimating Multimode fidelity using both extensive quantum state for balanced and unbalanced heterodyne detection across multiple sub-systems. Our multi-mode extension is based on the observation that even if for all the subsystems, the quantum states are close enough to the pure states; then we can conclude these quantum states are close to the tensor product of the pure states. As a result, it will produce a high fidelity witness. We construct a protocol for efficient boson sampling verification by using protocol-2 as a subroutine. In our quantum circuit, we demonstrated both unbalanced and balanced heterodyne detection. By choosing a threshold fidelity, we showed which fidelity will be used for further applications and which will be rejected. The verification gives rise to quantum speed-up.  \\

Lastly, we showed the use of heterodyne detection in a simple case of quantum key distribution and how it can be used for better distinction between the same and different encoding-decoding basis. We showed the use of heterodyne detection in one simple case. However, it can be extended to other cases as well which is left as an open problem. Quantum speed-up cannot be demonstrated in a single experiment, but rather in a series of tests. Our approaches will find a wide range of applications in current and future investigations for the trustworthy verification of quantum states.

\section{Acknowledgements} S.K.S., V.V. and S.P. would like to thank Bikash's Quantum (OPC) Pvt. Ltd. for providing hospitality during the course of this project. B.K.B. acknowledges IISER-K Institute fellowship. The authors also acknowledge Sayan Adhikari and Amlandeep Nayak of IISER Berhampur who helped us in plotting the graphs and peer review respectively. The authors appreciate IBM quantum experience's assistance in developing the fundamental circuits. The writers' opinions are their own and do not reflect IBM's or the IBM quantum experience team's official position.


\newpage
\begin{thebibliography}{100}
\expandafter\ifx\csname url\endcsname\relax
\def\url#1{\texttt{#1}}\fi
\expandafter\ifx\csname urlprefix\endcsname\relax\def\urlprefix{URL }\fi
\providecommand{\bibinfo}[2]{#2}
\providecommand{\eprint}[2][]{\url{#2}}

\bibitem{bs_quantadva}J. Preskill, Quantum computing and the entanglement frontier, arXiv:1203.5813 (2012).
\bibitem{bs_quantadva2}A. W. Harrow and A. Montanaro, Quantum computational supremacy, Nature \textbf{549}, 203 (2017).
\bibitem{bs_quantspeed}T. F. Ronnow, Z. Wang, J. Job, S. Boixo, S. V. Isakov, D. Wecker, J. M. Martinis, D. A. Lidar, and M. Troyer, Quantum computing: Defining and detecting quantum speedup, Science \textbf{345}, 420 (2014).
\bibitem{bs_shor}P. W. Shor, Polynomial-Time Algorithms for Prime Factorization and Discrete Logarithms on a Quantum Computer, SIAM Review \textbf{41}, 303 (1999).
\bibitem{bs_near}T. Haner, M. Roetteler, and K. M. Svore, Factoring using 2n+2 qubits with Toffoli based modular multiplication, Quantum Inf. Comput. \textbf{17}, pp (2017).

\bibitem{bs_mod}B. M. Terhal and D. P. DiVincenzo, Classical simulation of noninteracting-fermion quantum circuit, Phys. Rev. A, \textbf{65}, 032325 (2002).
\bibitem{bs_mod1}D. Shepherd and M. J. Bremner, Temporally unstructured quantum computation, Proc. R. Soc. A.,\textbf{465},1413–1439 (2009).
\bibitem{bs_mod2}M. J. Bremner, R. Jozsa, and D. J. Shepherd, Classical simulation of commuting quantum computations implies collapse of the polynomial hierarchy. Proc. R. Soc. A., \textbf{467}, 459 (2011).
\bibitem{bs_mod3}S. Boixo, S. V. Isakov, V. N. Smelyanskiy, Characterizing quantum supremacy in near-term devices, Nature Phys., \textbf{14}, 595–600 (2018). 
\bibitem{bs_mod4}R. Mezher, J. Ghalbouni, J. Dgheim, and D. Markham, Efficient approximate unitary t-designs from partially invertible universal sets and their application to quantum speedup, arXiv:1905.01504 (2019).
\bibitem{bs_demspeed}T. F. Ronnow, Z. Wang, et al. Quantum versus classical annealing of Ising spin glasses, Science, \textbf{345}, pp:420-424 (2014).
\bibitem{bs_submodel}A. Bouland, B. Fefferman, C. Nirkhe, et al., On the complexity and verification of quantum random circuit sampling, Nature Phys., \textbf{15}, 159–163 (2019). 
\bibitem{bs_submodel1}J. Eisert, D. Hangleiter, N. Walk, et al., Quantum certification and benchmarking, Nat. Rev. Phys., \textbf{2}, 382–390 (2020).
\bibitem{bs_bs}S. Aaronson and A. Arkhipov, Theory of Computing, \textbf{9}, 143 (2013).
\bibitem{bs_intro}B. T. Gard, K. R. Motes, J. P. Olson, P. P. Rhode and J. P. Dowling, An introduction to boson-sampling, arXiv:1406.6767 (2014).
\bibitem{bs_lininter}A. Crespi, R. Osellame, R. Ramponi, et al., Integrated multimode interferometers with arbitrary designs for photonic boson sampling, Nature Photon., \textbf{7}, 545–549 (2013). 
\bibitem{bs_classcomplex}D. Hangleiter, Sampling and the complexity of nature, arXiv:2012.07905 \textbf{[quant-ph]} (2020).
\bibitem{bs_molvib}J. Huh, G. Guerreschi, B. Peropadre, et al., Boson sampling for molecular vibronic spectra, Nature Photon., \textbf{9}, 615–620 (2015). 
\bibitem{bs_cry} G. M. Nikolopoulos, Cryptographic one-way function based on boson sampling, Quantum Inf. Process., \textbf{18}, 259 (2019).
\bibitem{bs_cry1}Z. Huang, P. P. Rhode, D. W. Berry, Photonic quantum data locking, Quantum, \textbf{5}, 447 (2021).
\bibitem{bs_cry2}G. M. Nikolopoulos and T. Brougham,  Decision and function problems based on boson sampling, Phys. Rev. A., \textbf{94}, 012315 (2016).
\bibitem{bs_matper}P. Clifford and R. Clifford, The classical complexity of Boson sampling, Proceedings of the ACM-SIAM Symposium on Discrete Algorithms, 146 (2018). 
\bibitem{bs_res}S. Goldstein, S. Korenblit, Y. Bendor, et al, Decoherence and interferometric sensitivity of boson sampling in superconducting resonator network, Phys. Rev. B ,\textbf{95}, 020502(R) (2017).
\bibitem{bs_photondem}J. B. Spring, B. J. Metcalf, P. C. Humpherys, et al., Boson Sampling on a Photonic Chip, Science \textbf{339}, 798-801 (2013).
\bibitem{bs_photodem1}F. Arute, K. Arya, R. Babbush, D. Bacon, et al., Quantum supremacy using a programmable superconducting processor, Nature, \textbf{574}, 505-510 (2019).
\bibitem{bs_photonvalid}N. Spagnolo, C. Vitelli, M. Bentivegna, et al., Experimental validation of photonic boson sampling, Nature Photon., \textbf{8}, 615–620 (2014). 
\bibitem{bs_photonvalid1}H. Wang, J. Qin, X. Ding, M. Chen, S. Chen, et al.,Boson Sampling with 20 Input Photons and a 60-Mode Interferometer in a $10^{14}$-Dimensional Hilbert Space, Phys. Rev. Lett., \textbf{123}, 250503 (2019).

\bibitem{bs_moldock}H. Zhong, L. Peng, Y. Li, Y. Hu, et al., Experimental Gaussian Boson sampling, Science Bulletin, \textbf{64}, 8 (2019)
\bibitem{bs_scatter}A. Lund, A. Laing, S. Rahimi-Keshari, et al., Boson sampling from a Gaussian state, Phys. Rev. Lett., \textbf{113}, 100502 (2014).
\bibitem{bs_gaussian}C. S. Hamilton, R. Kruse, L. Sansoni, S. Barkhofen, C. Silberhorn, I. Jex, Gaussian Boson Sampling, Phys. Rev. Lett., \textbf{119}, 170501 (2017).
\bibitem{bs_gaussian1}L. Chakhmakhchyan, N. Cerf, Boson sampling with Gaussian measurements, Phys. Rev. A., \textbf{96}, 032326 (2017).
\bibitem{bs_class}S. Rahimi-Keshari, A. Lund, T. Ralph, What can quantum optics say about computational complexity theory?, Phys. Rev. Lett., \textbf{114}, 060501 (2015).
\bibitem{bs_class1}S. Rahimi-Keshari, T. Ralph, C. Carlton, Efficient classical simulation of quantum optics, Phys. Rev X., \textbf{6 }, 021039 (2016).
\bibitem{bs_path}K. R. Motes, A. Gilchrist, J. P. Dowling, and P. P. Rohde, Scalable boson sampling with time-bin encoding using a loop-based architecture, Phys. Rev. Lett. \textbf{113}, 120501 (2014).
\bibitem{bs_path1} M. Pant and D. Englund, High dimensional unitary transformations and boson sampling on temporal modes using dispersive optics, Phys. Rev. A, \textbf{93}, 043803 (2016).
\bibitem{bs_crypto}D. Hangleiter, M. Kliesch, J. Eisert, and C. Gogolin, Sample Complexity of Device-Independently Certified “Quantum Supremacy”, Phys. Rev. Lett. \textbf{122}, 210502 (2019).
\bibitem{bs_stud}D. Shepherd and M. J. Bremner, Temporally unstructured quantum computation, Proc. R. Soc. A., \textbf{465}, 1413–1439 (2009).
\bibitem{bs_stud1}G. D. Kahanamoku-Meyer, Forging quantum data: classically defeating an IQP-based quantum test, arXiv:1912.05547 \textbf{[quant-ph]} (2019).
\bibitem{bs_stud2}MH Yung and B. Cheng, Deep learning of topological phase transitions from entanglement aspects: An unsupervised way, arXiv:2005.01510 \textbf{ [cond-mat.supr-con]} (2020).
\bibitem{bs_ass} U. Mahadev, Classical Verification of Quantum Computations, 2018 IEEE 59th Annual Symposium on Foundations of Computer Science (FOCS), pp. 259–267 (2018).
\bibitem{bs_ass1}Z. Brakerski, P. Christiano, U. Mahadev, U. Vazirani, and T. Vidick, A Cryptographic Test of Quantumness and Certifiable Randomness from a Single Quantum Device, 2018 IEEE 59th Annual Symposium on Foundations of Computer Science (FOCS), pp. 320–331 (2018).
\bibitem{bs_ass2}Z. Brakerski, V. Koppula, U. Vazirani, and T. Vidick, Simpler Proofs of Quantumness, arXiv:2005.04826 \textbf{[quant-ph]} (2020).
\bibitem{bs_valid}N. Spagnolo, C. Vitelli, M. Bentivegna, D. J. Brod, A. Crespi, F. Flamini, S. Giacomini, G. Milani, R. Ramponi, P. Mataloni, et al., Experimental validation of photonic boson sampling, Nature Photonics, \textbf{8}, 615-620 (2014).
\bibitem{bs_assu}S. Ferracin, T. Kapourniotis, and A. Datta, Accrediting outputs of noisy intermediate-scale quantum computing devices, New J. of Phys., \textbf{21}, 113038 (2019).
\bibitem{bs_proto}U. Chabaud, F. Grosshans, E. Kashefi, D. Markham, Efficient verification of Boson Sampling, arXiv:2006.03520 \textbf{ [quant-ph]} (2020).
\bibitem{bs_hetero}H. Yuen and J. Shapiro, Optical communication with two-photon coherent states--Part III: Quantum measurements realizable with photoemissive detectors, IEEE Transactions on Information Theory, \textbf{26}, 78-92 (1980).
\bibitem{bs_mfid}L. Aolita, C. Gogolin, M. Kliesch et al. Reliable quantum certification of photonic state preparations, Nat Commun \textbf{6}, 8498 (2015).
\bibitem{bs_inequality}C. A. Fuchs, J. V. D. Graaf, Cryptographic distinguishability measures
for quantum-mechanical states, arXiv:quant-ph/9712042 (1998)
\bibitem{bs_appbs}C. H. Bennett, G. Brassard, Quantum cryptography: Public key distribution and coin tossing, In Proceedings of IEEE International Conference on Computers, Systems and Signal Processing, \textbf{175}, 8, (1984).
\bibitem{bs_appbs1} P. W. Shor, J. Preskill, Simple Proof of Security of the BB84 Quantum Key Distribution Protocol, Phys. Rev. Lett., \textbf{85}, 441, (2000).
\end{thebibliography}
\end{document}